\documentclass[11pt,tightenlines,superscriptaddress,floatfixolumn,english,aps,notitlepage,nofootinbib]{revtex4-1}
\usepackage{xspace}
\usepackage[colorlinks=true,linkcolor=blue, citecolor=blue]{hyperref}

\usepackage{float}
\usepackage{graphicx}
\usepackage{epstopdf}
\usepackage{bbold}
\usepackage{comment}
\usepackage{color}
\usepackage{soul}
\usepackage[abs]{overpic}
\usepackage{amsmath}
\usepackage{lipsum}
\usepackage[abs]{overpic}
\usepackage{nicefrac}
\begin{document}
\title{Search for the chiral magnetic effect in collisions between two isobars with deformed and neutron-rich nuclear structures}
\author{Xin-Li Zhao}
\affiliation{Key Laboratory of Nuclear Physics and Ion-beam
  Application (MOE), Institute of Modern Physics, Fudan University,
  Shanghai 200433, China}
\affiliation{Shanghai Research Center for Theoretical Nuclear Physics, NSFC and Fudan University, Shanghai $200438$, China}
\author{Guo-Liang Ma}
\email[]{glma@fudan.edu.cn}
\affiliation{Key Laboratory of Nuclear Physics and Ion-beam
 Application (MOE), Institute of Modern Physics, Fudan University,
 Shanghai 200433, China}
\affiliation{Shanghai Research Center for Theoretical Nuclear Physics, NSFC and Fudan University, Shanghai $200438$, China}

\begin{abstract}
Isobar collisions which were thought to have the same background and different magnetic fields provide an opportunity to verify the chiral magnetic effect (CME) in relativistic heavy-ion collisions. However, the first result from the RHIC-STAR isobar experiment did not observe the CME signal, but discovered that the backgrounds are different between $_{44}^{96}\textrm{Ru}+_{44}^{96}\textrm{Ru}$ and $_{40}^{96}\textrm{Zr}+_{40}^{96}\textrm{Zr}$ collisions. We test eighteen cases of Woods-Saxon parameter settings resulting from different nuclear deformation or nuclear structure effects using a mutiphase transport model. We find out that seven cases can reasonably reproduce three reference ratios measured by the STAR experiment.  Considering both the halo-type neutron skin structure and CME-like charge separation, we demonstrate that it is difficult for the CME observables ($\Delta\delta$, $\Delta\delta$ ratio, $\Delta\gamma$ and $\Delta\gamma$ ratio) to distinguish the presence or absence of the CME, if the CME strength is weak in isobar collisions. It is because the final state interactions significantly weaken the initial CME signal, resulting in non-linear sensitivities of the CME observables to the CME strength. Therefore, more sensitive observables are required to search for the possible small CME signal in isobar collisions.

\end{abstract}

\maketitle

\section{Introduction}
\label{sec:introduction}
Deconfined QCD matter with quark and gluon degrees of freedom, the quark-gluon plasma (QGP), is believed to be created in the early stage of relativistic heavy-ion collisions at the Relativistic Heavy Ion Collider (RHIC)~\cite{STAR:2005gfr,PHENIX:2004vcz} and at the Large Hadron Collider (LHC)~\cite{ALICE:2008ngc}. In addition, some studies have shown that an extremely strong magnetic field can be generated in the early stage of QGP evolution~\cite{Skokov:2009qp,Bloczynski:2012en,Bzdak:2011yy,Deng:2012pc,Zhao:2017rpf,Zhao:2019ybo}. Thus relativistic collisions provide a suitable environment for some novel chiral anomalous phenomena, such as the chiral magnetic effect (CME)~\cite{Kharzeev:2004ey,Fukushima:2008xe,Kharzeev:2020jxw}, and the chiral magnetic wave (CMW)~\cite{Burnier:2011bf,Kharzeev:2015znc}. The efforts to search for these phenomena have been the focus of research for nearly a decade, because they may reveal new knowledge related to the local charge-parity violation in strong interactions and the topological fluctuations of the QCD vacuum. 

The researches of the CME have been in rapid development, although the backgrounds for seeking the CME cause serious difficulties~\cite{Bzdak:2012ia,Li:2020dwr,Wang:2016iov,Zhao:2019hta,Gao:2020vbh,Wang:2018ygc,Koch:2016pzl}. It was first proposed that the CME could be observed by measuring the $\gamma$ correlator~\cite{Voloshin:2004vk}. By assuming that the CME background is charge independent, the difference $\Delta\gamma$ between the opposite sign (OS) and the same sign (SS) is expected to reflect the CME signal. 
Positive $\Delta\gamma$ has been observed by both RHIC-STAR~\cite{STAR:2009wot,STAR:2009tro} and LHC-ALICE experiments~\cite{ALICE:2012nhw}, which is consistent with the CME expectation. Unfortunately, the real situation is complicated because some backgrounds can also cause a positive $\Delta\gamma$, such as resonance decays~\cite{Wang:2009kd} and local charge conservation~\cite{Schlichting:2010qia}, which makes it more difficult to observe a weak CME signal under a dominant background~\cite{Zhao:2018blc,STAR:2021pwb}. 
Therefore, the isobar program of $_{44}^{96}\textrm{Ru}+_{44}^{96}\textrm{Ru}$ and $_{40}^{96}\textrm{Zr}+_{40}^{96}\textrm{Zr}$ collision by the RHIC-STAR experiment was expected to solve this problem because the two isobar systems have a same nucleon number but different proton numbers~\cite{Voloshin:2010ut}. Isobar collisions are expected to produce the same background but different magnetic fields, which should make the CME observable ratio of $\rm Ru+Ru$ to $\rm Zr+Zr$ greater than unity~\cite{Xu:2017zcn,Deng:2018dut,Zhao:2019crj,Shi:2019wzi,Choudhury:2021jwd}. However, the recent blind analysis~\cite{STAR:2019bjg} carried by the five STAR experimental analysis groups has consistently shown that the ratios of the CME observables between the two isobar collisions are less than unity~\cite{STAR:2021mii}, which indicates that no predefined CME signal so far has been observed in the isobar program.

The backgrounds for the two isobar collisions are different in fact, since differences of charged-particle multiplicity and collective flow between the two isobar collisions have been observed by STAR~\cite{STAR:2021mii}. To explain the source of the differences, different nuclear deformations or nuclear structures have been used. As shown in Refs.~\cite{Xu:2021vpn,Xu:2021uar,Zhang:2021kxj} , the neutron-skin effect and quadruple and octupole deformations need to be taken into account to explain the measured differences, which supports the existence of neutron-skin structure and nuclear deformation in the two isobar nuclei~\cite{Centelles:2008vu,Butler:2016rmu,Cao:2020rgr,Chen:2021auq}. In other words, their nucleon density distributions should be described by a deformed Woods-Saxon (WS) form with some proper deformation parameters considering these effects. It has been suggested that the deviation of the ratio from unity has some origin in the nuclear structure, which can impact the initial state and survive to the final state~\cite{Jia:2021oyt}. It has been proposed that the isobar collision experiment provides a new experimental way to study nuclear deformation and nuclear structure~\cite{Giacalone:2021udy,Jia:2021tzt,Jia:2021qyu,Shuryak:2022ejk}. In this work, we first try to find the best setting of WS parameters which can reproduce the STAR isobar data, then constrain the strength of the possible CME signal in $\rm Ru+Ru$ and $\rm Zr+Zr$ collisions at $\sqrt{s}=$200 GeV by using a multiphase transport (AMPT) model including both the signal of the chiral magnetic effect and good setting of WS parameters.

The paper is organized as follows. First, we introduce and test different nuclear geometry configurations of isobar nuclei using the AMPT model to find which cases are good ones in  Sec.~\ref{sec:geometry}. Then we introduce the AMPT model including the CME and the CME observable of interest in Sec.~\ref{sec:correlator}. In Sec.~\ref{sec:results}, we present our results about the CME observable and compare them with the STAR data. Finally, we draw conclusions in Sec.~\ref{sec:summary}.

\section{Introducing geometry configurations of isobar nuclei}
\label{sec:geometry}
The AMPT model is a hybrid transport model which mainly includes four parts~\cite{Lin:2004en,Ma:2016fve,Lin:2021mdn}: the HIJING model for the initial condition, Zhang's parton cascade for the evolution of partonic stage, quark coalescence model for the hadronization,  and the ART model for hadron rescatterings. The nucleons inside colliding nuclei are initialized at the initial stage by the HIJING model~\cite{Wang:1991hta}.

To model nucleons inside $_{44}^{96}\textrm{Ru}$ and $_{40}^{96}\textrm{Zr}$, the spatial distribution of nucleons in the rest frame can be written in the WS form in spherical coordinates,
\begin{eqnarray}
\rho (r,\theta )&=&\rho _{0}/\{1+\rm{exp}[(r-R(\theta,\phi))/a]\}, \\
R(\theta ,\phi )&=&R_{0}[1+\beta_{2} Y_{2,0}(\theta,\phi)+\beta_{3} Y_{3,0}(\theta,\phi)],
\label{rho}
\end{eqnarray}%
where $\rho _{0}$ is the normal nuclear density, $R_{0}$ is the radius of the nucleus, $a$ is the surface diffuseness parameter, and $\beta _{2}$ and $\beta _{3}$ are the quadrupole and octupole deformities of the nucleus. We try to find all available settings of isobar geometry deformation parameters, as shown in Table~\ref{cases}. There are 15 cases of settings which will be tested in our work. 
The deformation parameter settings Case 1 and Case 2 are obtained from experimental measurement~\cite{Deng:2016knn,Fricke:1995zz,Raman:2001nnq,Moller:1993ed}. The old Case 1 and old Case 2~\cite{Zhao:2019crj} are actually from Case 1 and Case 2, but the corresponding parameters were tuned to ensure the consistency between the model simulation and the experimental measurement~\cite{Shou:2014eya}. Case 3, Cases 7-10, and skin-type and halo-type cases are taken from Xu's work~\cite{Xu:2021vpn} which considers the neutron-skin effect and/or a nuclear quadrupole deformity in the WS distribution with the TRENTO model~\cite{Moreland:2014oya}. Cases 4-6 are taken from Jia's recent researches~\cite{Jia:2021oyt,Zhang:2021kxj}, among which Case 4 and Case 5 can well reproduce some experimental results. Meanwhile, Case 6 is also from Ref.~\cite{Pritychenko:2013gwa}. Case 11 is taken from Ref.~\cite{Hammelmann:2019vwd} which introduced a deformed setting for Ru and a neutron-skin effect for Zr. Note that in the recent blind analysis, the STAR Collaboration has tabulated Case 1 to Case 3, and found that Case 3 is the best one of the Cases 1-3. In addition, we will also study Cases 1-3 with a fixed $\beta_{3}=0.235$. Therefore, we have eighteen cases of deformation settings in total.

\begin{table*}[]
\caption{The different cases of Woods-Saxon parameter settings tested in the AMPT model.}
\renewcommand\arraystretch{1.5}
\label{cases}

\begin{tabular}{cccc|ccc|ccc|ccc|ccc} \\
\hline\hline
\multicolumn{1}{l}{} & \multicolumn{3}{c|}{Case 1~\cite{Deng:2016knn,Fricke:1995zz,Raman:2001nnq}}   &\multicolumn{3}{c|}{Case 2~\cite{Deng:2016knn,Fricke:1995zz,Moller:1993ed}}  &\multicolumn{3}{c|}{Case 3~\cite{Xu:2021vpn}}  &\multicolumn{3}{c|}{Old Case 1~\cite{Zhao:2019crj,Shou:2014eya}}   &\multicolumn{3}{c}{Old Case 2~\cite{Zhao:2019crj,Shou:2014eya}} 
\\ \cline{2-16}
  &$R_{0}$&a& $\beta_{2}$    &$R_{0}$&a&$\beta_{2}$    &$R_{0}$&a&$\beta_{2}$    &$R_{0}$&a&$\beta_{2}$    &$R_{0}$&a& $\beta_{2}$       
  \\  \hline
$_{44}^{96}$Ru &5.085&0.46&0.158   & 5.085&0.46&0.053   &5.067&0.500&0.00   &5.13&0.46&0.13   &5.13&0.46&0.03\\ 
$_{40}^{96}$Zr  &5.020&0.46&0.080   &5.020 &0.46&0.217   &4.965&0.556&0.00  &5.06&0.46&0.06   &5.06&0.46&0.18\\    
 \hline  \hline \\
\end{tabular}

\begin{tabular}{ccccc|cccc|cccc|ccc|ccc}
\hline\hline
\multicolumn{1}{l}{} & \multicolumn{4}{c|}{Case 4~\cite{Jia:2021oyt}}   &\multicolumn{4}{c|}{Case 5~\cite{Jia:2021oyt}}  &\multicolumn{4}{c|}{Case 6~\cite{Zhang:2021kxj,Pritychenko:2013gwa}}  &\multicolumn{3}{c|}{Case 7~\cite{Xu:2021vpn}}  &\multicolumn{3}{c}{Case 8~\cite{Xu:2021vpn}}  
\\ \cline{2-19}
  &$R_{0}$&a& $\beta_{2}$& $\beta_{3}$    &$R_{0}$&a&$\beta_{2}$& $\beta_{3}$    &$R_{0}$&a&$\beta_{2}$& $\beta_{3}$   &$R_{0}$&a&$\beta_{2}$   &$R_{0}$&a&$\beta_{2}$      
  \\  \hline
$_{44}^{96}$Ru &5.09&0.46&0.162&0   &5.09&0.46&0.162&0    &5.09&0.52&0.154&0  &5.065&0.485&0.16  &5.085&0.523&0  \\ 
$_{40}^{96}$Zr  &5.09&0.52&0.060&0.2   &5.02&0.46&0.060&0.2   &5.09&0.52&0.060&0.2  &4.961&0.544&0.16  &5.021&0.523&0  \\    
 \hline  \hline \\
\end{tabular}

\begin{tabular}{ccccc|ccc|ccc|ccc|ccc}
\hline\hline
\multicolumn{2}{l}{}  &\multicolumn{3}{c|}{Case 9~\cite{Xu:2021vpn}}   &\multicolumn{3}{c|}{Case 10~\cite{Xu:2021vpn}}   &\multicolumn{3}{c|}{Case 11~\cite{Hammelmann:2019vwd}}    &\multicolumn{3}{c|}{Skin-type~\cite{Xu:2021vpn}}  &\multicolumn{3}{c}{Halo-type~\cite{Xu:2021vpn}}   
\\ \cline{3-17}
 \multicolumn{2}{l}{}  &$R_{0}$&a& $\beta_{2}$    &$R_{0}$&a& $\beta_{2}$    &$R_{0}$&a& $\beta_{2}$    &$R_{0}$&a&$\beta_{2}$   &$R_{0}$&a&$\beta_{2}$           
  \\  \hline
$_{44}^{96}$Ru &n   & 5.075&0.505&0  & 5.073&0.490&0.16  &5.085&0.46&0.158  &5.085&0.523&0   &5.085&0.523&0     \\ 
$_{44}^{96}$Ru &p   & 5.060&0.493&0  & 5.053&0.480&0.16  &5.085&0.46&0.158  &5.085&0.523&0   &5.085&0.523&0     \\ 
$_{40}^{96}$Zr  &n   & 5.015&0.574&0  & 5.007&0.564&0.16  &5.080&0.46&0    &5.194&0.523&0   &5.021&0.592&0     \\ 
$_{40}^{96}$Zr  &p   & 4.915&0.521&0  & 4.912&0.508&0.16  &5.080&0.34&0    &5.021&0.523&0   &5.021&0.523&0     \\    
 \hline  \hline
\end{tabular}
\end{table*}

\begin{figure*}[htb]
\includegraphics[width=1.\textwidth]{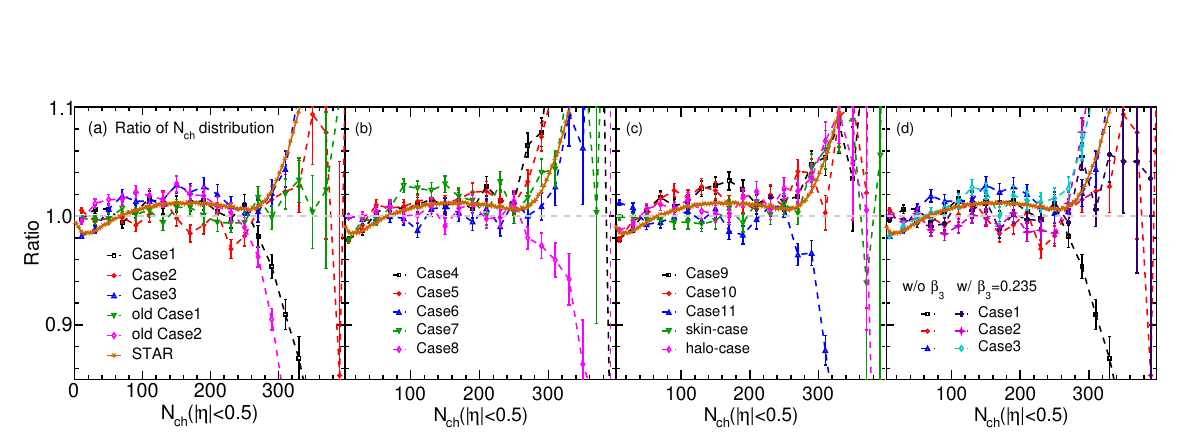}
\caption{The ratios of the charged-particle multiplicity distribution ($|\eta|<0.5$) in $\rm Ru+Ru$ collisions to that in $\rm Zr+Zr$ collisions from the AMPT model using eighteen cases of Woods-Saxon parameter settings, in comparison with the experimental data~\cite{STAR:2021mii}.}
\label{fig:ratio1}
\end{figure*}

\begin{figure*}[htb]
\includegraphics[width=1.\textwidth]{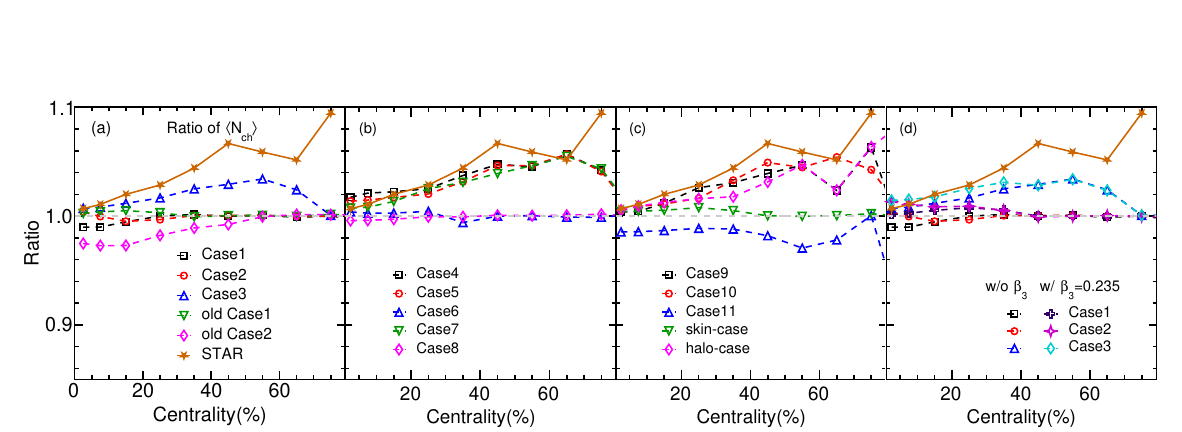}
\caption{The centrality dependences of the ratio of the average number of charged particles ($|\eta|<0.5$) in $\rm Ru+Ru$ collisions to that in $\rm Zr+Zr$ collisions from the AMPT model using eighteen cases of Woods-Saxon parameter settings, in comparison with the experimental data~\cite{STAR:2021mii}.}
\label{fig:ratio2}
\end{figure*}

\begin{figure*}[htb]
\includegraphics[width=1.\textwidth]{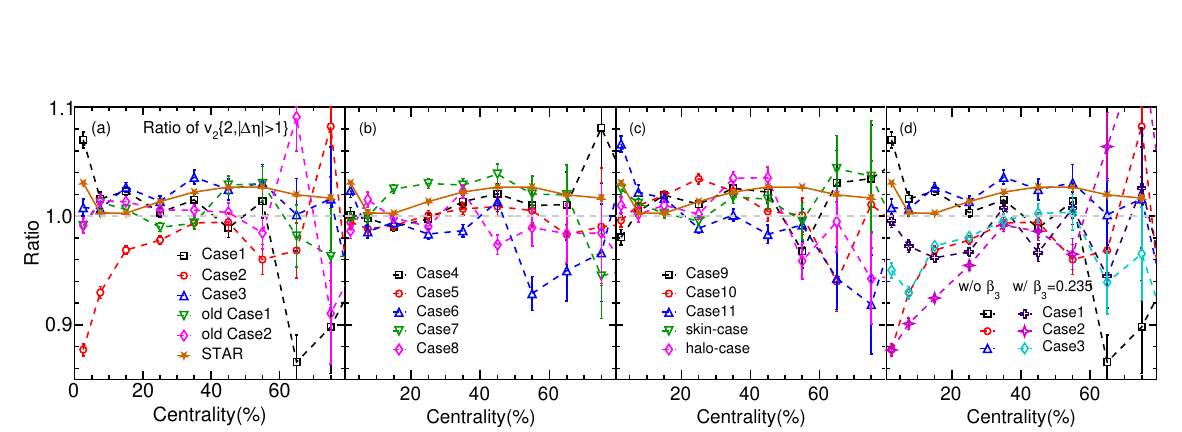}
\caption{The centrality dependences of the ratio of $v_{2}\{2,|\Delta\eta|>1\}$) in $\rm Ru+Ru$ collisions to that in $\rm Zr+Zr$ collisions from the AMPT model using eighteen cases of Woods-Saxon parameter settings, in comparison with the experimental data~\cite{STAR:2021mii}.}
\label{fig:ratio3}
\end{figure*}

In order to determine which case(s) is (are) good geometry setting(s) for the two isobar nuclei, we will test if the above cases of WS parameters can produce reasonable ratios of Ru to Zr in the charged-particle multiplicity distribution, average number of charged particles and elliptic flow $v_{2}$, as shown in Figs.~\ref{fig:ratio1}-\ref{fig:ratio3} ($10^{6}$ minimum-bias events for each case), respectively. 
Figure~\ref{fig:ratio1} shows the ratios of the charged-particle multiplicity ($N_{\rm ch}$) distribution in $\rm Ru+Ru$ collisions to that in $\rm Zr+Zr$ collisions from eighteen cases of WS parameters compared to the STAR data. It is obvious that the ratios from Case 1, old Case 2, Case 8, and Case 11 are less than unity for high $N_{\rm ch}$, different from the experimental result which is above 1. For the other cases, they can basically reproduce the experimental data, i.e. the ratio is close to unity for low $N_{\rm ch}$, but greater than unity for high $N_{\rm ch}$ .
In addition, Fig.~\ref{fig:ratio1}(d) shows the results from Cases 1-3 but with an octupole deformation parameter of $\beta_{3}$=0.235 for Zr. We see that with the octupole deformation, the ratio of the $N_{\rm ch}$ distribution gets larger than unity for high $N_{\rm ch}$ for the Case 1. This indicates that the octupole deformation for Zr decreases the yield of charged particles in central $\rm Zr+Zr$ collisions. However, the Cases 2 and 3 are less affected by the octupole deformation.
Overall, the old Case 1, Case 1 with $\beta_{3}$, Cases 2-3 without and with $\beta_{3}$, Cases 4-7, Cases 9-10, the skin-type case, and the halo-type case can reproduce the STAR measured ratio of charged-particle multiplicity distribution.

Next we check if they can reproduce the centrality dependence of the ratio of the average number of charged particles $\langle N_{\rm ch} \rangle$ in $\rm Ru+Ru$ collisions to that in $\rm Zr+Zr$ collisions. As done by STAR, we use the multiplicity distribution of all charged particles within the pseudorapidity window of $|\eta|<0.5$ to define centrality bins. 
In Fig.~\ref{fig:ratio2}(a)-(c), we can see that the ratios in Case 3, Case 4, Case 5, Case 7, Case 9, Case 10, and the halo-type case increase from central to peripheral centrality bins, consistent with the STAR measurement. Considering the additional effect of octupole deformation in Cases 1-3, Fig.~\ref{fig:ratio2}(d) shows that the ratio will change slightly for central collisions but remain unchanged for peripheral collisions. Overall, Case 3 without and with $\beta_{3}$, Cases 4-5, Case 7, Cases 9-10,  halo-type case can reproduce the centrality dependence of the ratio of average number of charged particles between the two isobar collisions measured by STAR.

Last, we check if these cases can reproduce the centrality dependence of the ratio of elliptic flow in $\rm Ru+Ru$ collisions to that in $\rm Zr+Zr$ collisions. 
As done by STAR, we use the two-particle correlation method to calculate the elliptic flow $v_{2}\{2\}$ and the charged particles under kinetic cuts of $|\eta|<1$ and $0.2<p_{\rm T}<2~\textrm{GeV}/c$. 
From Figs.~\ref{fig:ratio3}(a)-(c), we find that the ratios of $v_{2}\{2,|\Delta\eta|>1\}$ from the Cases 3-5, Case 7, Cases 9-10, and the skin-type and halo-type cases can describe the experimental data. In terms of the effect of octupole deformation, Fig.~\ref{fig:ratio3}(d) shows that the finite $\beta_{3}$ decreases the ratio of $v_{2}\{2,|\Delta\eta|>1\}$ especially for central collisions.

According to the above performance results of all cases, we can see Cases 3-5, Case 7, Cases 9-10 and the halo-type case are good cases, if we define the criterion for a good case as whether it can qualitatively reproduce the above three kinds of ratios at the same time. It is worth noting that Cases 4-6 and the effect of $\beta_{3}$ have been studied in detail in Refs.~\cite{Jia:2021oyt,Jia:2021qyu,Zhang:2021kxj}, and their recent study suggested that the ratios calculated at the same $N_{\rm ch}$ provide a better baseline for the non-CME background~\cite{Jia:2022iji}. To find which case better describes the data other than by visual judgment, the chi-square ($\chi^{2}$) test is utilized to compare the three ratios of the model results to the experimental data. 
As shown in Table~\ref{tab:chi2}, we calculate the relative mean $\chi^{2}$ according to the three observables in Figs.\ref{fig:ratio1}-\ref{fig:ratio3}.
We find that the halo-type neutron skin case is the best with a minimum value of $\chi^{2}$ among the eighteen cases. Thus, we will choose the halo-type neutron skin case for our following study related to the CME. 

\begin{table}[]
\caption{The relative mean $\chi^{2}$ for the eighteen cases of Woods-Saxon parameter settings.}
\label{tab:chi2}
\setlength{\tabcolsep}{3mm}{
\begin{tabular}{ccccccc}
 \hline  \hline
     &Old Case 1 &Old Case 2    & Case 1  & Case 2  & Case 3 & Case 4              \\ \hline
$\left <\chi^{2} \right>$  & 0.204      &0.682 & 0.255   & 0.400   & 0.097     & 0.053  \\          
\\ 
     & Case 5   & Case 6    & Case 7 & Case 8     & Case 9        & Case 10 \\  \hline
$\left <\chi^{2} \right>$  & 0.049          & 0.224           & 0.051  & 0.227      & 0.057    & 0.048 \\ 
\\
     & Case 11 &   Skin-type      & $\textbf{Halo-type}$   & Case 1 with $\beta_{3}$  & Case 2 with $\beta_{3}$ & Case 3 with $\beta_{3}$  \\    \hline
$\left <\chi^{2} \right>$   & 0.430   & 0.177   & $\textbf{0.047}$    & 0.247          & 0.506           & 0.166          \\
 \hline  \hline
\end{tabular} }
\end{table}

\section{Introducing and observing the CME}
\label{sec:correlator}
Previously, we introduced a CME-like charge separation into the initial stage of the parton cascade in the AMPT model, the details of which are in Ref.~\cite{Ma:2011uma}. By tuning the percentage $f$ of how many quarks join the charge separation,\footnote{Note that because the AMPT model includes only the degrees of freedom for quarks, $f$ is an approximate percentage.} the signal strength of the CME can be controlled in the AMPT model. The percentage $f$ is defined as,
\begin{equation}
f = \frac{N_{\uparrow(\downarrow)}^{+(-)}-N_{\downarrow(\uparrow)}^{+(-)}}{N_{\uparrow(\downarrow)}^{+(-)}+N_{\downarrow(\uparrow)}^{+(-)}},
 \label{eq-f}
\end{equation}
where $N$ is the number of quarks of a given species ($N_{f}$=3), $+$ and $-$ denote positive and negative charges, and $\uparrow$ and $\downarrow$ represent the moving directions along the magnetic field. 
We realistically make the initial charge separation according to both the magnitude and direction of the magnetic field by calculating the magnetic field for each event. Based on our previous work~\cite{Zhao:2019crj}, $f$ for $\rm Ru+Ru$ collisions is 1.15 times larger than $\rm Zr+Zr$ collisions. We require that $f_{\rm Ru+Ru}/f_{\rm Zr+Zr}=1.15$ for all centrality bins, for example, when we show $f=10\%$, it means $f_{\rm Ru+Ru}=10\%$ and $f_{\rm Zr+Zr}=10\%/1.15=8.7\%$.

To observe the CME, we choose the differences $\Delta\gamma$ and $\Delta\delta$ between two correlators with opposite sign and the same sign to observe the CME signal, where the $\gamma$ correlator $\gamma_{\alpha \beta } =\left<\rm cos(\phi_{\alpha}+\phi_{\beta}-2\Psi_{RP} )\right>$ and $\delta_{\alpha \beta }=\left<cos(\phi_{\alpha}-\phi_{\beta})\right>$, where $\phi_{\alpha,\beta}$ are the azimuthal angles of charged particles and $\Psi_{\rm RP} $ is the reaction plane angle~\cite{STAR:2009tro,ALICE:2012nhw,Bzdak:2010fd}. A more positive $\Delta\gamma$ or more negative $\Delta\delta$ is expected to correspond to a stronger CME, as expected by the CME~\cite{Shi:2019wzi}. To cancel out the background contribution, we further obtain the observable of the $\Delta\gamma$ ratio between two isobar collisions to compare with the STAR measurement.

\section{Results and discussions}
\label{sec:results}

\begin{figure*}[htb]
\includegraphics[width=0.9\textwidth]{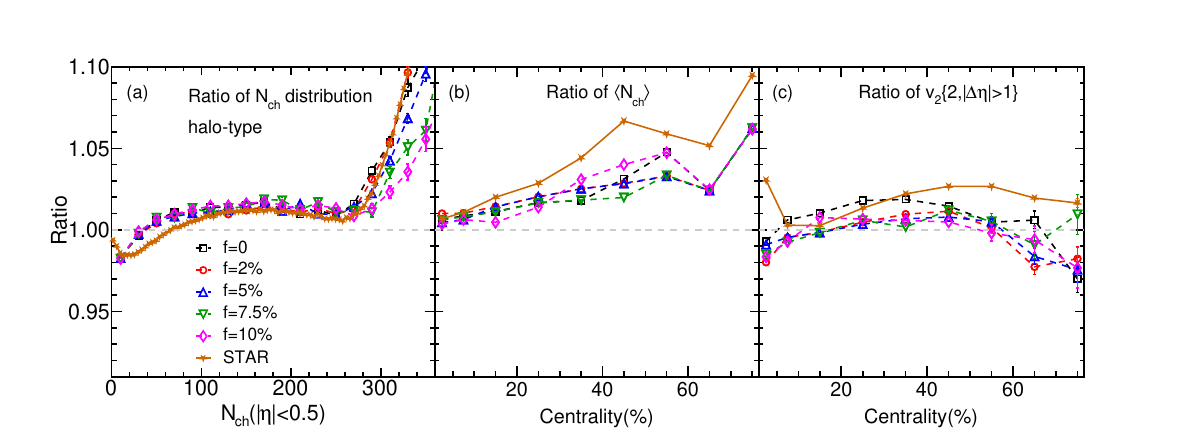}
\caption{The ratio of the charged-particle multiplicity distribution (a), the ratio of the average number of charged particles (b) and the ratio of $v_{2}\{2,|\Delta\eta|>1\}$ (c) in $\rm Ru+Ru$ collisions to that in $\rm Zr+Zr$ collisions from the AMPT model with different strengths of the CME, in comparison with the experimental data~\cite{STAR:2021mii}.}
\label{fig:ratio4}
\end{figure*}

\begin{figure}[htb]
\includegraphics[width=0.65\textwidth]{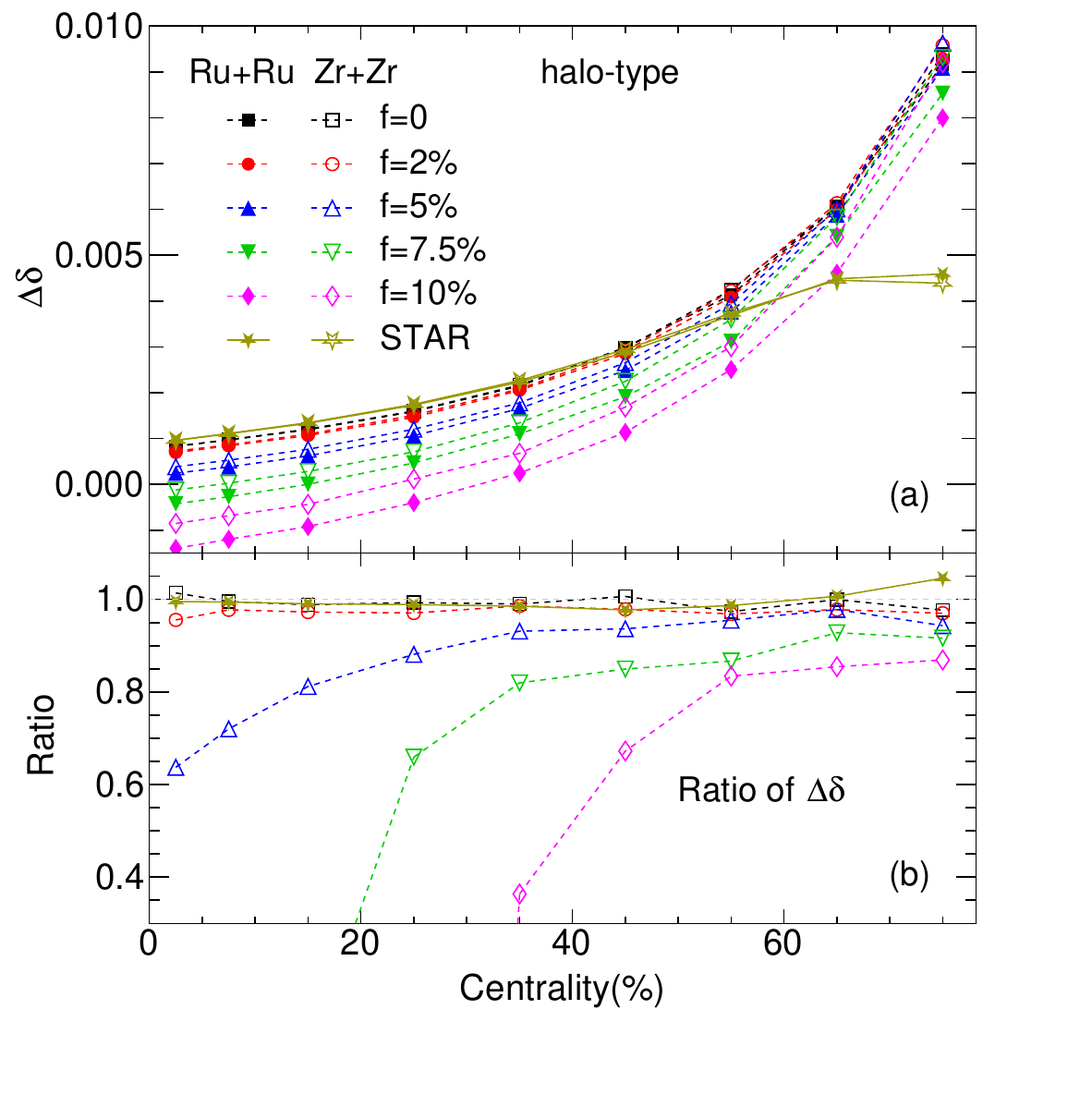}
\caption{The centrality dependences of $\Delta\delta$ (a) and the ratio of $\Delta\delta$ in $\rm Ru+Ru$ collisions to that in $\rm Zr+Zr$ collisions (b) from the AMPT model with different strengths of the CME, in comparison with the experimental data~\cite{STAR:2021mii}.}
\label{fig:ratiodelta}
\end{figure}

\begin{figure}[htb]
\includegraphics[width=0.65\textwidth]{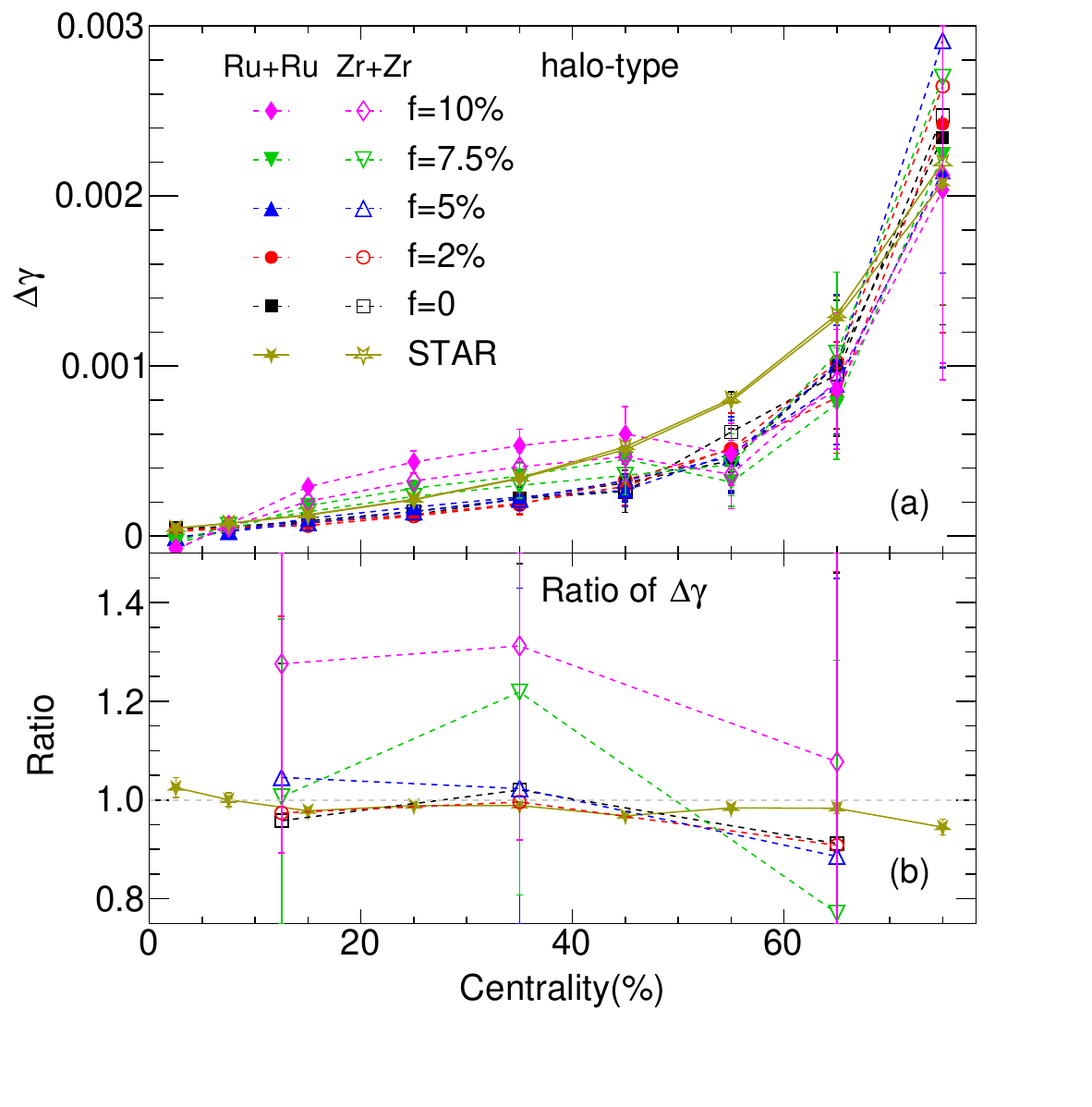}
\caption{The centrality dependences of $\Delta\gamma$ (a) and the ratio of $\Delta\gamma$ in $\rm Ru+Ru$ collisions to that in $\rm Zr+Zr$ collisions (b) from the AMPT model with different strengths of the CME, in comparison with the experimental data~\cite{STAR:2021mii}.}
\label{fig:ratiodr}
\end{figure}

\begin{figure}[htb]
\includegraphics[width=0.65\textwidth]{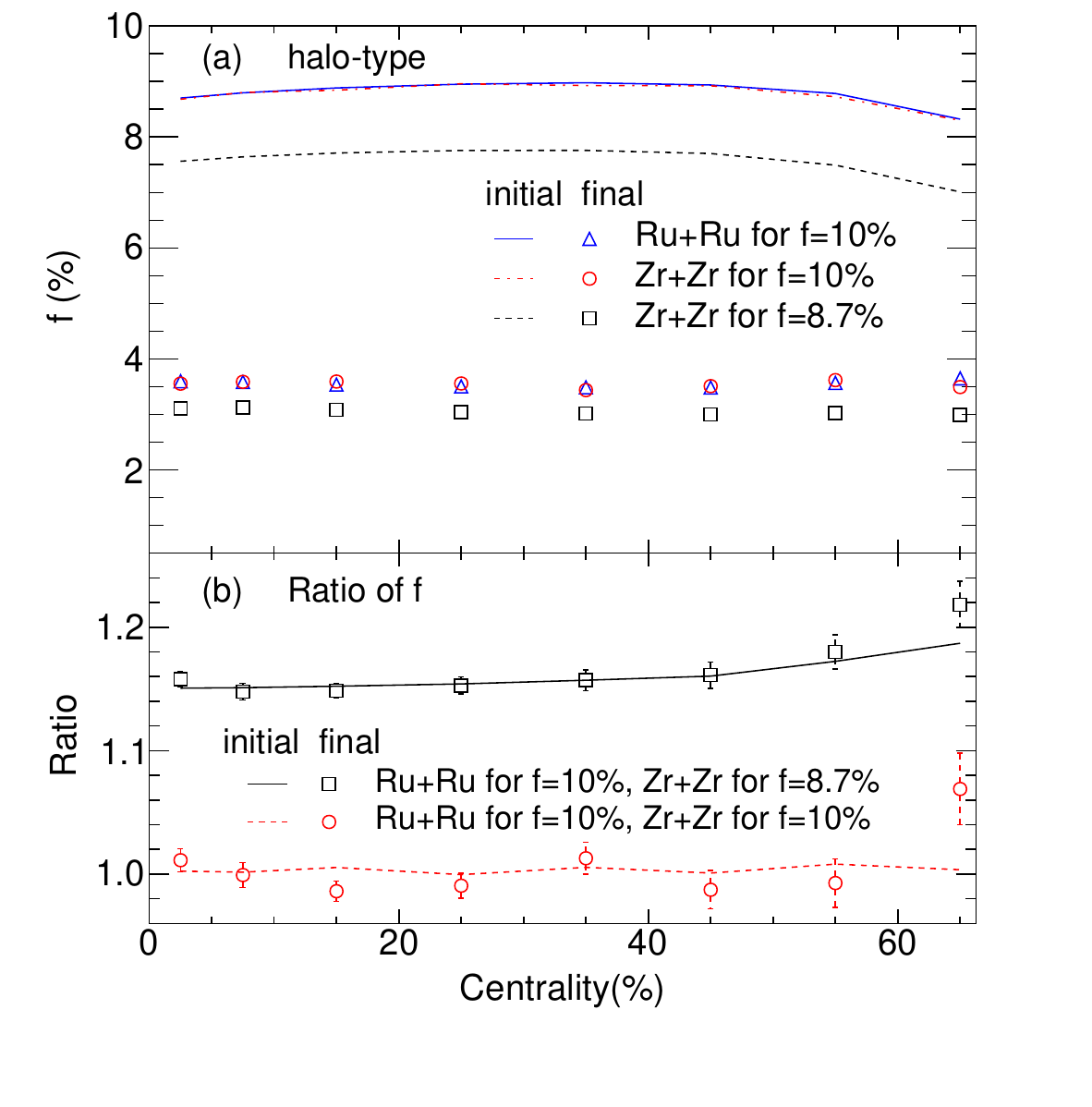}
\caption{The centrality dependences of charge separation percentage (a) and the ratios of charge separation percentage in $\rm Ru+Ru$ collisions to that in $\rm Zr+Zr$ collisions (b) at the initial stage (lines) and final stage (symbols) from the AMPT model with different initial charge separation percentages $f$.}
\label{fig:ratiocme}
\end{figure}

Before studying the CME observables, we need to make sure the ratios of the charged-particle distribution and $v_{2}\{2,|\Delta\eta|>1\}$ have no big change if the CME signal is introduced into the AMPT model. Figure~\ref{fig:ratio4} shows the above three criterion ratios for the AMPT model without the CME and with different strengths of the CME, in comparison with the experimental data. All three ratios from the AMPT model with different strengths of the CME are close to those from the AMPT model without the CME, and all are consistent with the experimental data. This indicates that the introduction of the initial CME signal into collisions between two isobars with halo-type neutron skin structures has little effect on the three reference ratios. Therefore, the AMPT model with the halo-type of Woods-Saxon parameter setting provides a good baseline to study the CME in isobar collisions.

The centrality dependences of $\Delta\delta$ for the two isobar collisions from the AMPT model without the CME and with different strengths of the CME are shown in Fig.~\ref{fig:ratiodelta}(a), and compared with the experimental data. For central and mid-central collisions, the AMPT model without the CME or with a small strength of the CME can reproduce the experimental data, but results are  lower than the data if the CME strength is larger than $f=5\%$, because the CME tends to reduce $\Delta\delta$. However, all the AMPT results fail to describe $\Delta\delta$ for the most peripheral collisions. The centrality dependences of the ratio of $\Delta\delta$ for the two isobar collisions are presented in Fig.~\ref{fig:ratiodelta}(b). The AMPT results without the CME and a CME strength of $f=2\%$ are closer to the data than those with larger CME strengths, indicating that the CME strength is very small, or absent, in isobar collisions.

The centrality dependences of $\Delta\gamma$ for the two isobar collisions from the AMPT model without the CME and with different strengths of the CME are shown in Fig.~\ref{fig:ratiodr}(a), and compared with the experimental data. We find that the AMPT results are above the experimental data if the strength of the CME is $f=7.5\%$ or $f=10\%$ for mid-central collisions. 
However, the results for $f=2\%$ or $f=5\%$ are almost as same as that without the CME ($f=0$), all of which can well describe the experimental data. This is consistent with the previous findings that the CME observable $\gamma$ has a nonlinear sensitivity to the strength of the CME signal, as the final state interactions significantly weaken the initial CME signal~\cite{Deng:2018dut,Ma:2011uma,Huang:2019vfy}, which will be discussed later.

Then the centrality dependences of the ratio of $\Delta\gamma$ between two isobar collisions are presented in Fig.~\ref{fig:ratiodr}(b). The ratios for $f=7.5\%$ and $10\%$ look too large to describe the experiment data, especially for the centrality bin of $20-50\%$. Compared to the experimental data, no CME or a small CME strength is favored in order to reproduce the data. However, it is difficult to distinguish no CME from a small strength of the CME, because the ratios from $f=0$, $2\%$ and $5\%$ agree with each other in our limited statistics ($\approx 3\times10^{7}$  minimum-bias events for each case, while about $4\times10^{9}$ events are used by STAR). This indicates that the initial CME signal is absent or small in isobar collisions. Considering $\Delta\delta$ and $\Delta\gamma$ in Figs.~\ref{fig:ratiodelta} and \ref{fig:ratiodr}, our findings are consistent with the recent simulations from event-by-event anomalous-viscous fluid dynamics (EBE-AVFD), which suggests that the STAR results imply a limited CME signal contribution of about 2\%~\cite{Kharzeev:2022hqz}.
However, due to our limited number of events, the statistical errors in the AMPT simulations are large. Compared to the error in the STAR data, we estimate that the AMPT simulation requires at least about $1.8\times10^{9}$ events to distinguish such a small CME intensity. Since it is quite computationally expensive, we leave it for further study in the future.

According to Eq.~(\ref{eq-f}), the charge separation percentages for $\rm Ru+Ru$ and $\rm Zr+Zr$ can be calculated. Figure~\ref{fig:ratiocme}(a) illustrates the charge separation percentages as a function of centrality bin at the initial stage and final stage in AMPT model. Note that the calculated $f$ is not exactly equal to the input parameter $f$, because the imported initial charge separation is determined by both the magnitude and direction of magnetic field, i.e., $\langle B^{2} \rm cos~2(\Psi_{B}-\Psi_{RP}) \rangle$. It is obvious that compared to the initial stage (lines), the $f$ at the final stage (symbols) significantly decreases for the two isobar collisions. This indicates that the final state interactions dramatically weaken the initial CME signal. In Fig.~\ref{fig:ratiocme}(b), we first check that the ratio of $f$ between the two isobar collisions is consistent with one if the initial CME strengths of the two isobar collisions are as the same as for $f$=10\%. On the other hand, the ratio of $f$ between the two isobar collisions at the final stage is consistent with that at the initial stage, which is consistent with our previous study~\cite{Deng:2018dut}. It is a good news for searching for the CME, because the CME difference between the two isobaric collisions with the halo-type setting can be reserved until the final stage. On the other hand, it indicates that the observable $\Delta\gamma$ is not sensitive to a small strength of the CME. Therefore, more sensitive observables are required to search for possible small CME signals in isobar collisions.

\section{Conclusions}
\label{sec:summary}
We tested eighteen cases of Woods-Saxon parameter settings which consider either nuclear deformation or nuclear neutron-skin effect for $\rm Ru+Ru$ and $\rm Zr+Zr$ collisions at $\sqrt{s}=$200 GeV, using the AMPT model. Only seven of the eighteen cases (Case 3, Case 4, Case 5, Case 7, Case 9, Case 10, and the halo-type case) can reasonably reproduce the experimental ratios of charged-particle multiplicity distribution, average number of charged particles, and elliptic flow, which demonstrates that the nuclear deformation and structure information have a non-negligible impact. Isobar collisions can serve for further research of nuclear deformation or nuclear neutron-skin structure, which currently has important implications for both nuclear structure and nuclear astrophysics~\cite{Lattimer:2004pg,Li:2008gp,Tsang:2012se,Li:2019xxz}.

Utilizing the chi-square ($\chi^{2}$) test,
we choose the halo-type case to study the CME using the AMPT model with different strengths of the CME. The measured $\Delta\delta$, $\Delta\gamma$, $\Delta\delta$ ratio, and $\Delta\gamma$ ratio can be reproduced by the AMPT model without the CME or with a small strength of the CME. On the other hand, they can not be described by the AMPT model with a stronger strength of the CME. This indicates that the initial CME signal in isobar collisions is absent or small in isobar collisions. It is due to the fact that the final state interactions significantly weaken the initial CME signal, resulting in the non-linear sensitivity of the CME observables. Therefore, more sensitive observables are required for searching for the possible small CME signal in isobar collisions. 

\begin{acknowledgments}
We thank Dr. Chun-Jian Zhang for helpful discussions and Dr. Chen Zhong for maintaining the high-quality performance of the Fudan supercomputing platform for nuclear physics. This work is supported by the National Natural Science Foundation of China under Grant No. 12105054, the China Postdoctoral Science Foundation under Grant No. 2021M690708 (X.Z.), the National Natural Science Foundation of China under Grants No.12147101, No. 11961131011, No. 11890710, No. 11890714, and No. 11835002, the Strategic Priority Research Program of the Chinese Academy of Sciences under Grant No. XDB34030000, and the Guangdong Major Project of Basic and Applied Basic Research under Grant No. 2020B0301030008 (G.M.).

\end{acknowledgments}

\end{document}